\magnification\magstephalf
\overfullrule 0pt
\input epsf
\font\rfont=cmr10 at 10 true pt
\def\ref#1{$^{\hbox{\rfont {[#1]}}}$}


\font\fourteenbf=cmbx12 scaled\magstep1

\font\eightrm=cmr8


\def\pmb#1{\setbox0=\hbox{#1}
 \kern.05em\copy0\kern-\wd0 \kern-.025em\raise.0433em\box0 }

\def \half {{\scriptstyle {1 \over 2}}}

 %


\def\boxit#1{\vbox{\hrule\hbox{\vrule\kern1pt\vbox
{\kern1pt#1\kern1pt}\kern1pt\vrule}\hrule}}

\def\h{\hfill\break}
\parskip=6pt
\parindent=0pt
\hsize=17truecm\hoffset=-5truemm
\vsize=24.5truecm
\def\footnoterule{\kern-3pt
\hrule width 17truecm \kern 2.6pt}


\catcode`\@=11 

\def\nolabels{\def\wrlabeL##1{}\def\eqlabeL##1{}\def\reflabeL##1{}}
\def\writelabels{\def\wrlabeL##1{\leavevmode\vadjust{\rlap{\smash%
{\line{{\escapechar=` \hfill\rlap{\sevenrm\hskip.03in\string##1}}}}}}}%
\def\eqlabeL##1{{\escapechar-1\rlap{\sevenrm\hskip.05in\string##1}}}%
\def\reflabeL##1{\noexpand\llap{\noexpand\sevenrm\string\string\string##1}}}
\nolabels
\global\newcount\refno \global\refno=1
\newwrite\rfile
\def\defref{$^{{\hbox{\rfont [\the\refno]}}}$\nref}
\def\nref#1{\xdef#1{\the\refno}\writedef{#1\leftbracket#1}%
\ifnum\refno=1\immediate\openout\rfile=refs.tmp\fi
\global\advance\refno by1\chardef\wfile=\rfile\immediate
\write\rfile{\noexpand\item{#1\ }\reflabeL{#1\hskip.31in}\pctsign}\findarg}
\def\findarg#1#{\begingroup\obeylines\newlinechar=`\^^M\pass@rg}
{\obeylines\gdef\pass@rg#1{\writ@line\relax #1^^M\hbox{}^^M}%
\gdef\writ@line#1^^M{\expandafter\toks0\expandafter{\striprel@x #1}%
\edef\next{\the\toks0}\ifx\next\em@rk\let\next=\endgroup\else\ifx\next\empty%
\else\immediate\write\wfile{\the\toks0}\fi\let\next=\writ@line\fi\next\relax}}
\def\striprel@x#1{} \def\em@rk{\hbox{}} 
\def\lref{\begingroup\obeylines\lr@f}
\def\lr@f#1#2{\gdef#1{\defref#1{#2}}\endgroup\unskip}
\newwrite\lfile
{\escapechar-1\xdef\pctsign{\string\%}\xdef\leftbracket{\string\{}
\xdef\rightbracket{\string\}}}

\def\writestop{\def\writestoppt{\immediate\write\lfile{\string\p
ageno%
\the\pageno\string\startrefs\leftbracket\the\refno\rightbracket%
\string\def\string\secsym\leftbracket\secsym\rightbracket%
\string\secno\the\secno\string\meqno\the\meqno}\immediate\closeout\lfile}}
\def\writestoppt{}\def\writedef#1{}
\catcode`\@=12 
\rightline{   DAMTP-2008-3 \ \ \ \      MAN/HEP/2008/1}\h

\centerline{\fourteenbf Successful description of exclusive vector meson electroproduction}
\bigskip
\bigskip
\centerline{A Donnachie}

\centerline{Department of Theoretical Physics}

\centerline{University of Manchester}

\centerline{\eightrm sandy@hep.man.ac.uk}
\bigskip
\centerline{P V Landshoff}

\centerline{Department of Applied Mathematics and Theoretical Physics}

\centerline{University of Cambridge}

\centerline{\eightrm pvl@damtp.cam.ac.uk}
\vskip 10truemm
{\bf Abstract}

Data for the differential cross sections for $\gamma^* p\to\rho^0 p$, $\gamma^*
p \to \phi p$ and $\gamma^* p\to J/\psi p$, at all available values of $Q^2$,  
fit very well to a combination of soft pomeron, hard pomeron and $f_2$ 
exchange.

\vskip 5truemm

In a previous paper\defref\oldrho{
A Donnachie and P V Landshoff, Physics Letters B478 (2000) 146
}, 
we presented a zero-parameter fit to data for the differential 
cross section $d\sigma /dt$
for exclusive $\rho^0$ photoproduction (that is, from real photons)
and pointed out that its success was a
triumph for Regge theory. In order to achieve this zero-parameter fit,
we made two simple assumptions: that it is valid to use $\rho^0$
dominance, and that the differential cross section for $\rho^0 p$
scattering is identical to that for $\pi^0 p$ scattering. While it
was surprisingly successful to make these assumptions, there is no
reason to suppose that they are valid exactly, and so the appearance
of new and more accurate data from the H1 collaboration\defref\hone{
H1 collaboration: J Olsson, 14th International Workshop on Deep Inelastic 
Scattering , Tsukuba, JAPAN (2006); H1 prelim-06-011
},
albeit preliminary, call for a reassessment.

The Regge approach to inelastic lepton scattering is steadily gaining 
support\defref\caldwell{
A Caldwell, arXiv:0802.0769
}.
In this paper, we successfully apply it to differential-%
cross-section data for exclusive $\rho$ production for $2.8 \leq W 
\leq 71.7$ GeV and to the ZEUS data\defref\zeus{
ZEUS collaboration: J Breitweg et al,  European Physical Journal C2 (1998) 
247 and C14 (2000) 213; S Checkanov et al, PMC Physics A1 (2007) 6
} 
for exclusive $\rho$ electroproduction which go up to $Q^2$ values of more than
40~GeV$^2$. We then go on to consider the ZEUS data\defref\phidat
{ZEUS collaboration: S Chekanov et al, Nuclear Physics B718 (2005) 3
} 
for the differential cross section for exclusive $\phi$ electroproduction, 
and the H1 and ZEUS 
data\defref\zeusjpsi{ZEUS collaboration: S Chekanov et al, European Physical Journal 
C 24 (2002) 345
}\defref\h1jpsi{H1 collaboration: A. Aktas et al, European Physical Journal 
C46 (2006) 585
}
for exclusive $J/\psi$ production.
\bigskip
{\bf Real-photon exclusive \pmb{$\rho$} production}

The H1 data start at values of $W$ greater than 20~GeV and, at small $t$,
show $d\sigma/dt$ rising steadily with increasing $W$, as is characteristic
of pomeron exchange. However, as is seen in figure 1, the
old fixed-target data\defref\fixed{
J Ballam et al, Physical Review D7 (1973) 3150;
Eisenberg WIS-71-9-PH (1971);
CLAS collaboration: M Battaglieri et al, Physical Review Letters 87 (2001) 
172002;
Omega Photon collaboration: M Aston et al, Nuclear Physics B209 (1982) 56
}
at lower values of $W$ show that $d\sigma/dt$ at each fixed value of $t$
initially decreases with increasing $W$,
indicating the presence of a significant contribution from $f_2,a_2$ exchange. 
It is therefore necessary to include these fixed-target data in any fit.
This we will do, along with the data from ZEUS.
\topinsert
\centerline{\epsfxsize=0.4\hsize\epsfbox[176 555 440 760]{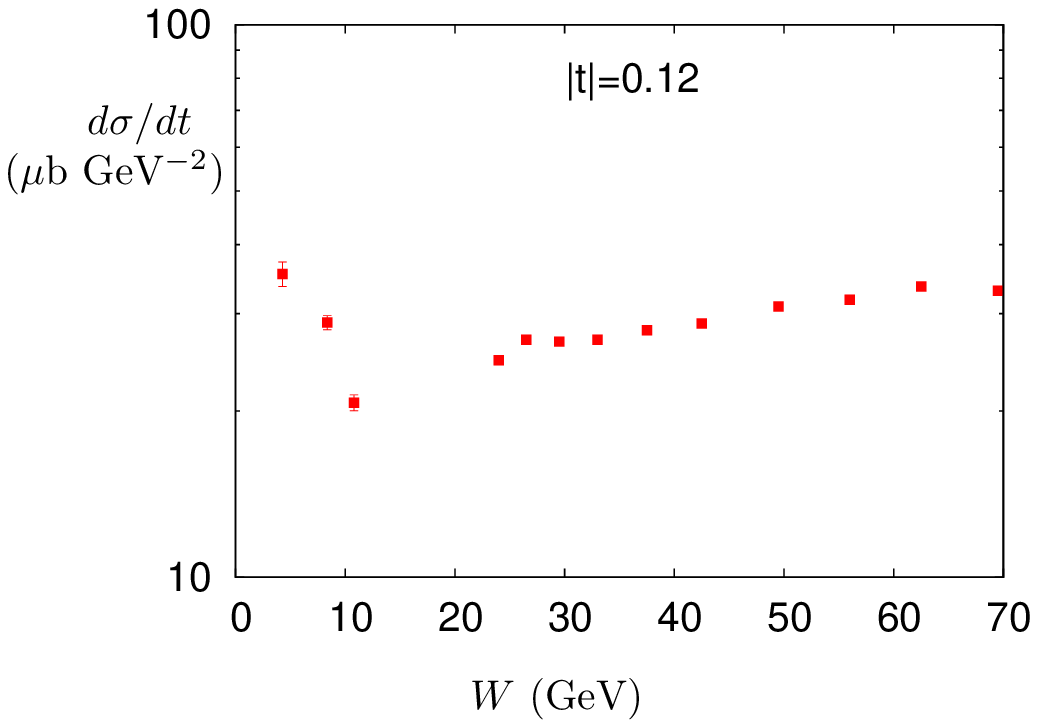}}

\centerline{Figure 1: Energy dependence of data for exclusive $\rho$ 
photoproduction at a fixed value of $t$.}
\endinsert
For hadronic elastic scattering $AB\to AB$, the high-energy data fit very well 
to\defref\book{
Sandy Donnachie, G\"unter Dosch, Peter Landshoff and Otto Nachtmann,
{\sl Pomeron physics and QCD}, Cambridge University Press (2002)
}
an amplitude whose $C=+$ form is
$$
T(s,t)=i\sum_i~X_i~F_A(t)F_B(t)~e^{-\half i\pi \alpha_i(t)}~(\nu/\nu _i^0)^{\alpha_i(t)-1}~~~~~~~~~~~~~\nu=\half (s-u)
\eqno(1)
$$
Here, the sum ranges over the exchanges of the soft pomeron and the
$f_2$ and $a_2$. The contribution from the $C=-$ exchanges $\rho,\omega$ is similar, except
that the phase is different\ref{\book}. The functions $\alpha(t)$ are the Regge
trajectories corresponding to the exchanges; for the soft pomeron we
have taken\ref{\book}
$$
\alpha_1(t)=1.08+\alpha'_1 t~~~~~~~~~\alpha'_1=0.25\hbox{ GeV}^{-2}
\eqno(2)
$$
and for $\rho,\omega,f_2,a_2$
$$
\alpha_2(t)=0.45+\alpha'_2 t~~~~~~~~~\alpha' _2=0.93\hbox{ GeV}^{-2}
\eqno(2b)
$$
and we have suggested that it is correct to choose $\nu _i^0=\half
{\alpha^\prime_i}^{-1}$
 ($i=1,2,3$) as in the Veneziano model.
The $X_i$ are real constants, and  $F_A(t)$ and $F_B(t)$ are the 
elastic form factors of the hadrons $A$ and $B$, in the case of the
proton its Dirac form factor
$$
F_1(t)={4m_p^2-2.79t\over 4m_p^2-t}{1\over (1-t/0.71)^2}
\eqno(3a)
$$

We use a form similar to (1) for the process $\gamma p\to\rho p$, 
but of course $\rho$ and $\omega$ exchange do not contribute, and also
we include an additional term, corresponding to hard-pomeron
exchange. Such a term is clearly seen in data for deep inelastic
electron scattering\defref\deepinel{
A Donnachie and P V Landshoff, Physics Letters B518 (2001) 63
}
and it may even be present in hadron-hadron scattering\defref\factor{
A Donnachie and P V Landshoff, Physics Letters B595 (2004) 393
}. That will surely be the case if it turns out that the correct
Tevatron total cross section is that quoted by CDF\defref\cdf{
CDF collaboration: F Abe et al, Physical Review D50 (1994)5550
}
rather than E710\defref\amos{
E710 collaboration: N A Amos et al, Physical Review Letters 63 (1989) 2784
}. 
We believe that the hard pomeron couples to the short-distance structure
of the initial hadrons. The short-distance component of the photon is 
more prominent than that of the proton and we have already noted\ref{\oldrho} 
that the ZEUS data for $\gamma p\to\rho p$ suggest a need for a hard-pomeron
contribution.

For the elastic form factor $F_A(t)$ in (1) we again use the Dirac
form factor of the proton. In place of $F_B(t)$ we need the
$\gamma\to\rho$ transition form factor. Previously\ref{\oldrho} we
took the pion form factor for this:
$$
G(t)={1\over 1-t/M^2}
\eqno(3b)
$$
with $M^2=0.5$ GeV$^2$. In our new fit, we allow $M$ to be a free parameter.
The other free parameters in our fit are the couplings $X_0,X_1,X_2$
respectively of hard-pomeron, soft-pomeron and $f_2,a_2$ exchange. We 
take\ref{\deepinel} the
intercept $1+\epsilon_0$ of the hard-pomeron trajectory to be
1.44. We do not take
the slope $\alpha'_0$ of the hard-pomeron trajectory to be a free parameter,
as it turns out that the quality of the fit is not very sensitive to it.
We have previously 
concluded\ref{\oldrho} that preliminary data for $\gamma p\to J/\psi~p~$
require a rather small value, and the fit we describe below reaffirms this, 
so we choose $\alpha'_0=0.01$~GeV$^{-2}$. 

\pageinsert
\line{\epsfxsize=0.4\hsize\epsfbox[176 555 440 760]{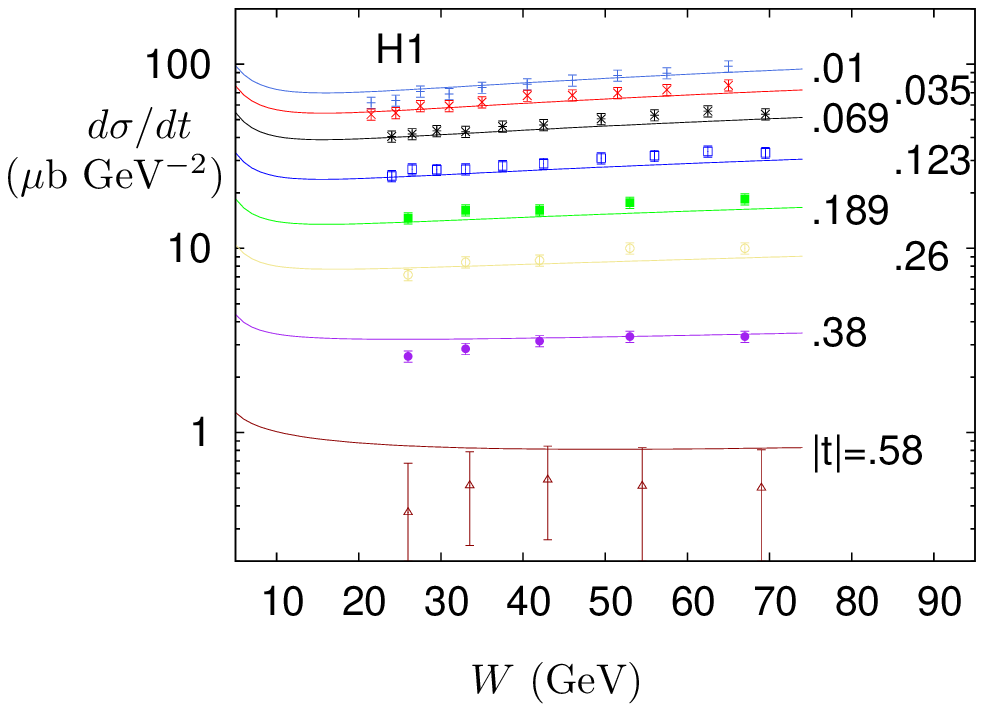}
\hfill
\epsfxsize=0.4\hsize\epsfbox[176 555 440 760]{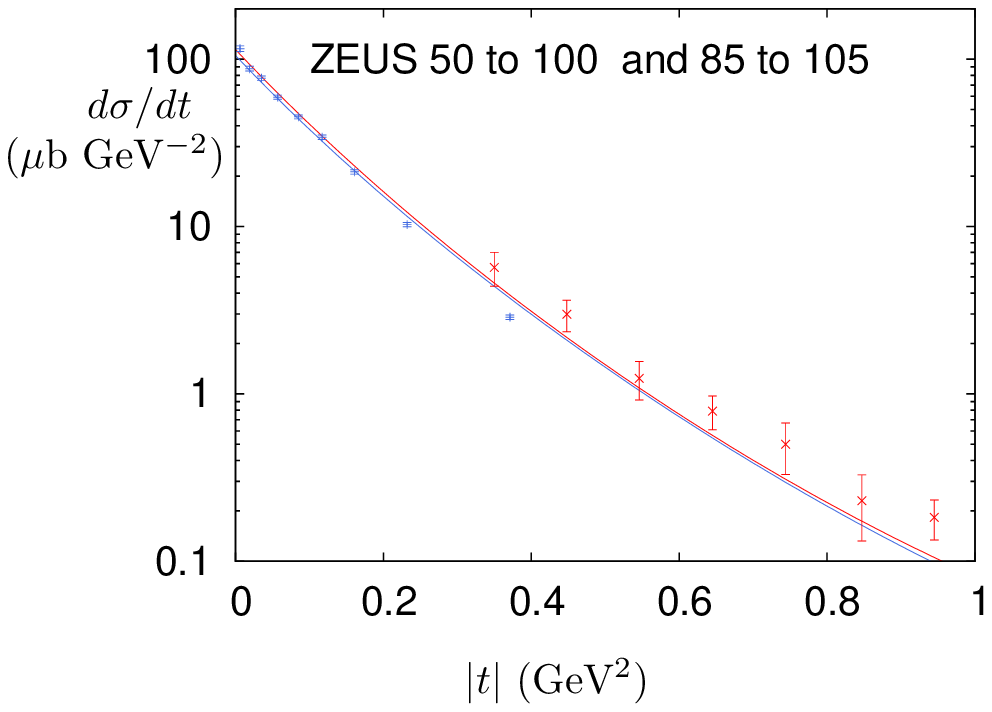}}
\medskip
\line{\epsfxsize=0.4\hsize\epsfbox[176 555 440 760]{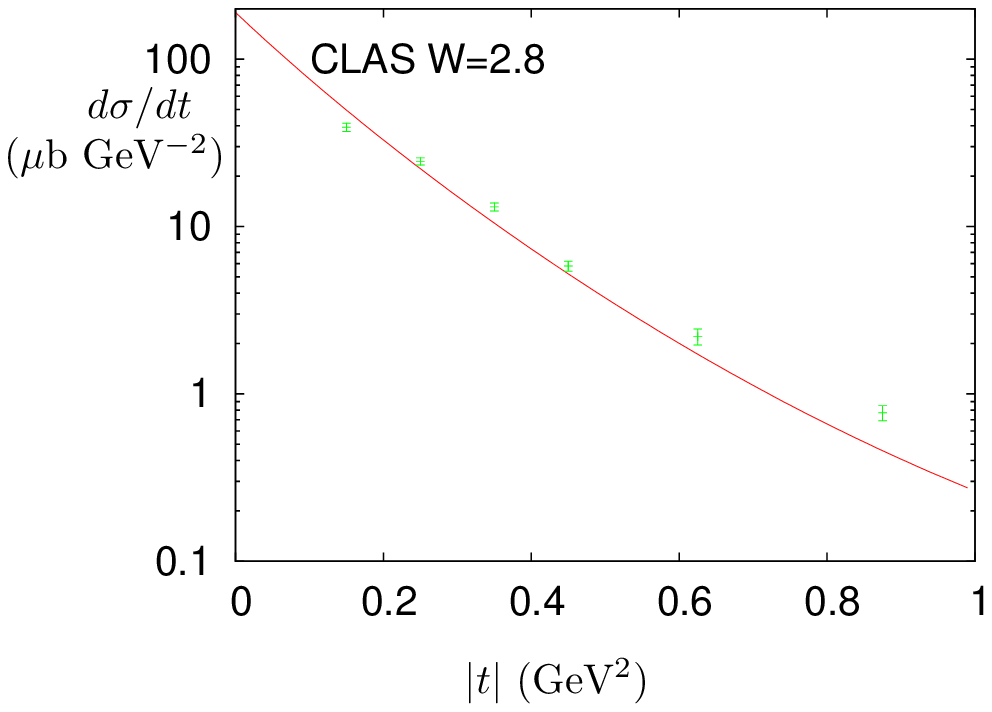}
\hfill
\epsfxsize=0.4\hsize\epsfbox[176 555 440 760]{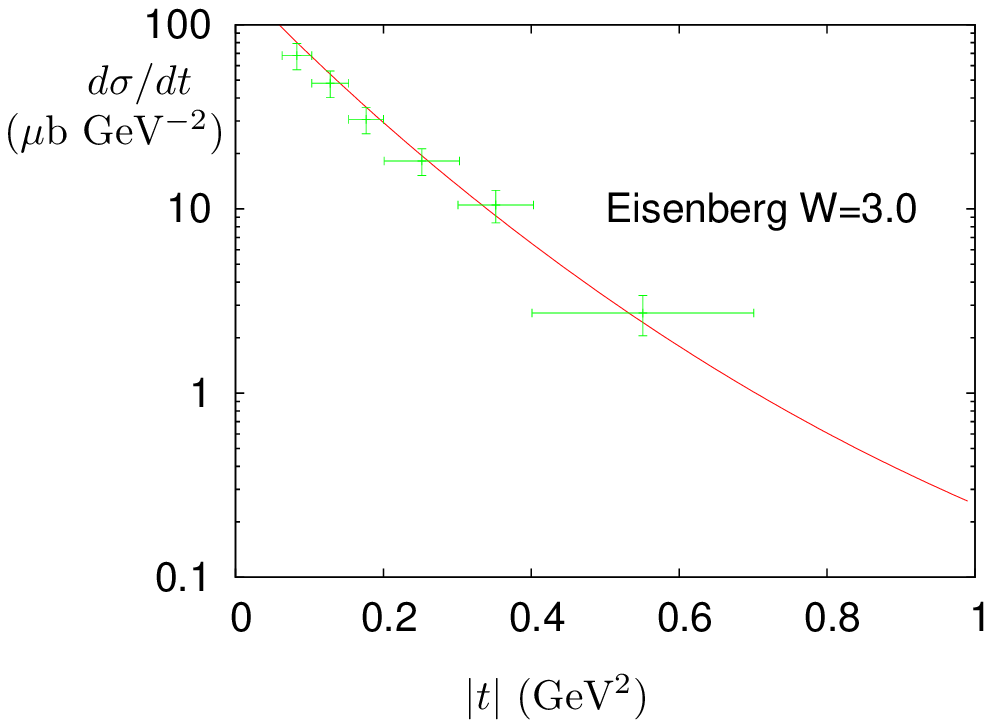}}
\medskip
\centerline{\epsfxsize=0.4\hsize\epsfbox[176 555 440 760]{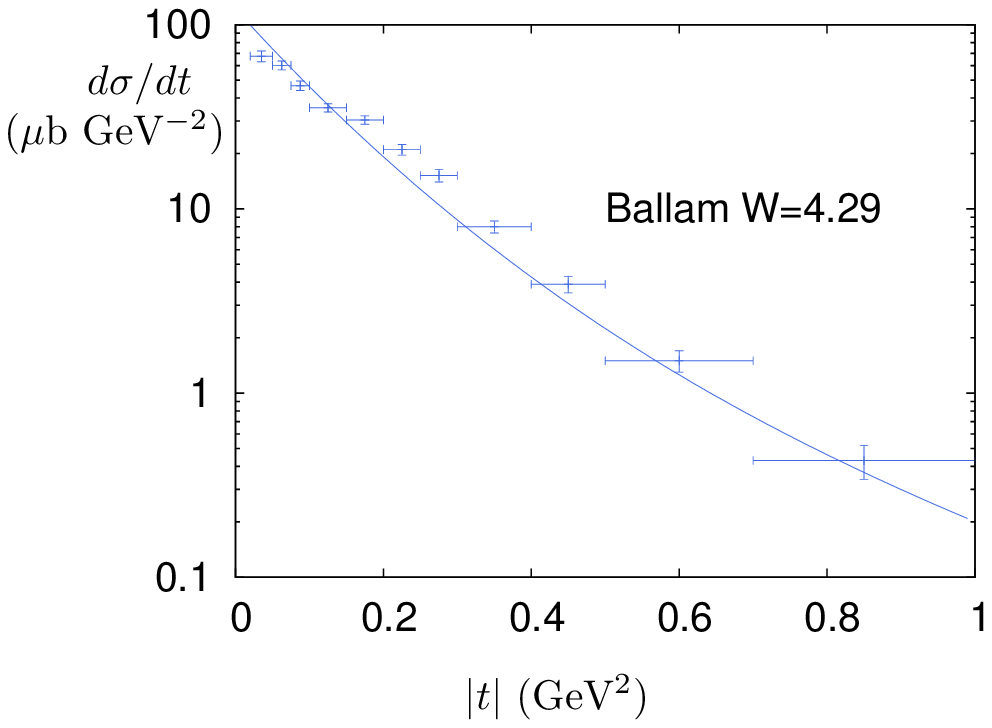}}
\medskip
\line{\epsfxsize=0.4\hsize\epsfbox[176 555 440 760]{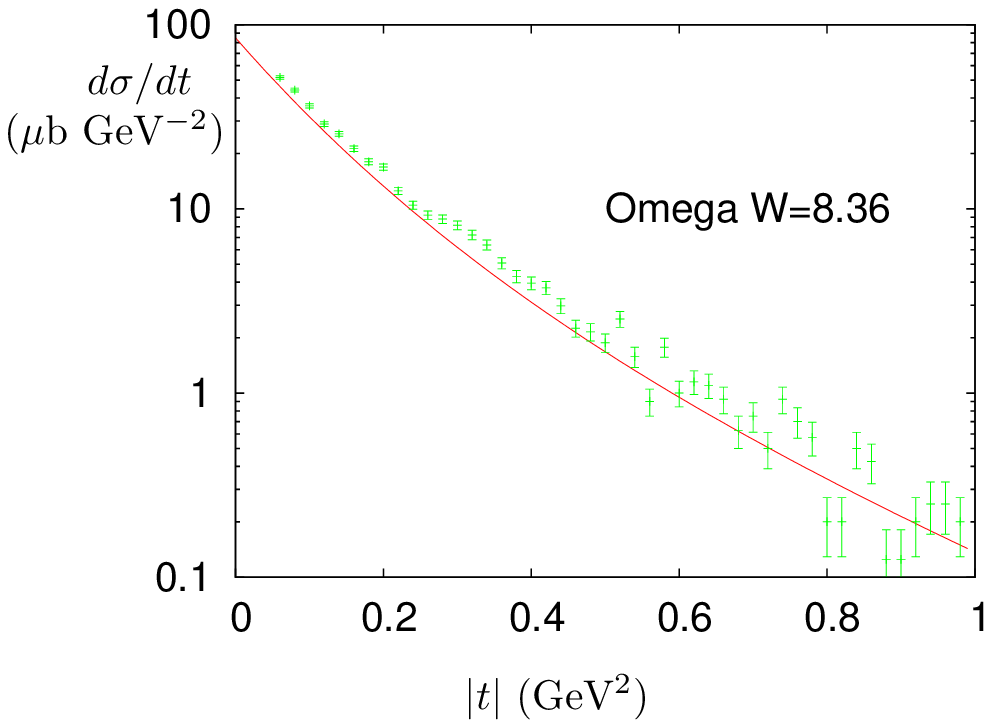}
\hfill
\epsfxsize=0.4\hsize\epsfbox[176 555 440 760]{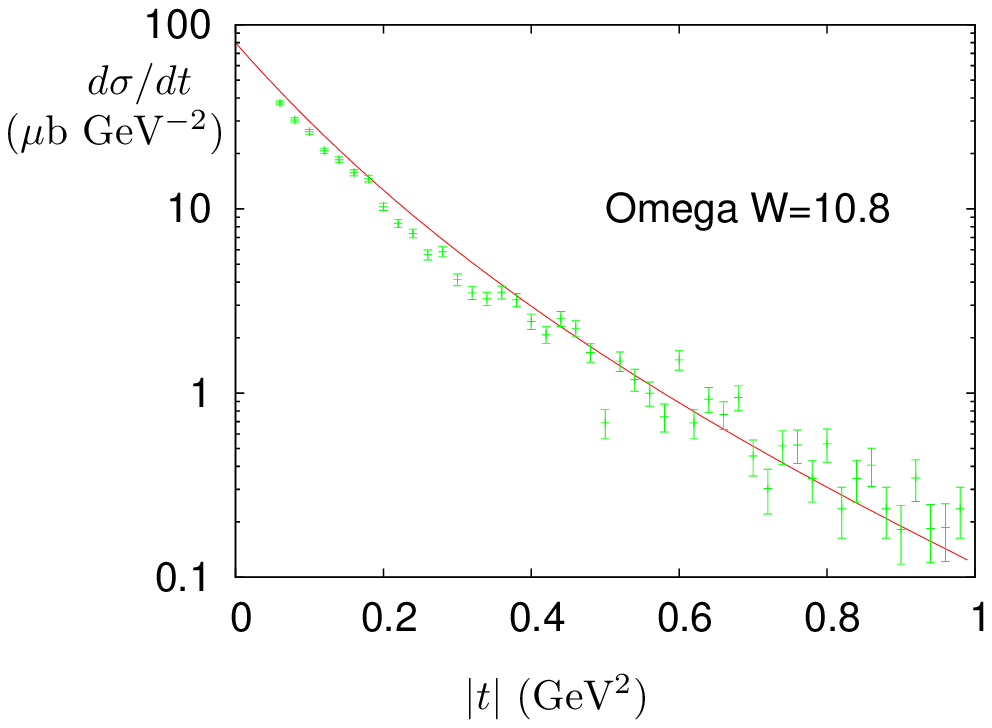}}
\medskip
\centerline{Figure 2: Fits to data for exclusive $\rho$ 
photoproduction}
\endinsert

The fit then results in
$$
X_0=0.069~~~~ X_1= 5.33~~~~ X_2=21.1~~~~ M=0.91~\hbox{GeV}
\eqno(4)
$$
where the units for the $X_i$ are such that $|T(s,t)|^2$ is the differential cross
section in $\mu$b GeV$^{-2}$.  The comparison with the high-energy
data is shown in the upper plots of figure~2. The ZEUS data were taken over two energy ranges
and the figures suggest that there may be a problem with the higher-energy
data.
The fits to the low-energy data are shown in the lower plots of figure 2. 
The Omega data have been normalised from their fit to the un-normalised 
differential cross section and their quoted total cross section.

Because of the small value of $X_0$, the contribution from the hard pomeron
to the fit is small and, except at the largest values of $W$ and $t$, the
hard-pomeron contribution is not really required. 
\bigskip
{\bf Exclusive \pmb{$\rho$} electroproduction}

For non-zero $Q^2$, the differential cross section receives contributions
from both transverse and longitudinal photons and its form is not well
constrained by our understanding of QCD. So our fit is largely informed
by trial and error.

An obvious guess is to make the replacement 
$$
M^2 \to M^2+Q^2
\eqno(5a)
$$
in (3b). Because the $f_2,a_2$ exchange contribution is small at the energies
for which there are data, we omit it. Initially the coefficients of the soft
and hard pomerons were fitted to the data at each $Q^2$ to give an indication 
of their $Q^2$ dependence. A good description of the $Q^2$ dependence of $X_0$ 
and $X_1$ is obtained if we multiply them by
$$
H(Q^2;Q^2_{1i},Q^2_{2i},Q^2_{3i})= 
{1\over (1+Q^2/Q^2_{1i})^2}+{Q^2/Q^2_{2i}\over (1+Q^2/Q^2_{3i})^3}
~~~~~~i=0,1
\eqno(5b)
$$
So, keeping $X_0,X_1$ and $M$ fixed, we make our fit varying the 6 parameters  
in (5b). This results in the
values
$$
Q_{10}^2=8.2~~~Q_{20}^2=3.9~~~Q_{30}^2=0.16
$$$$
Q_{11}^2=1.8~~~Q_{21}^2=63.6~~~Q_{31}^2=10.7
\eqno(5c)
$$
all in GeV$^2$ units.  The fit is shown in 
figure~3.
\topinsert
\line{\epsfxsize=0.4\hsize\epsfbox[176 555 440 760]{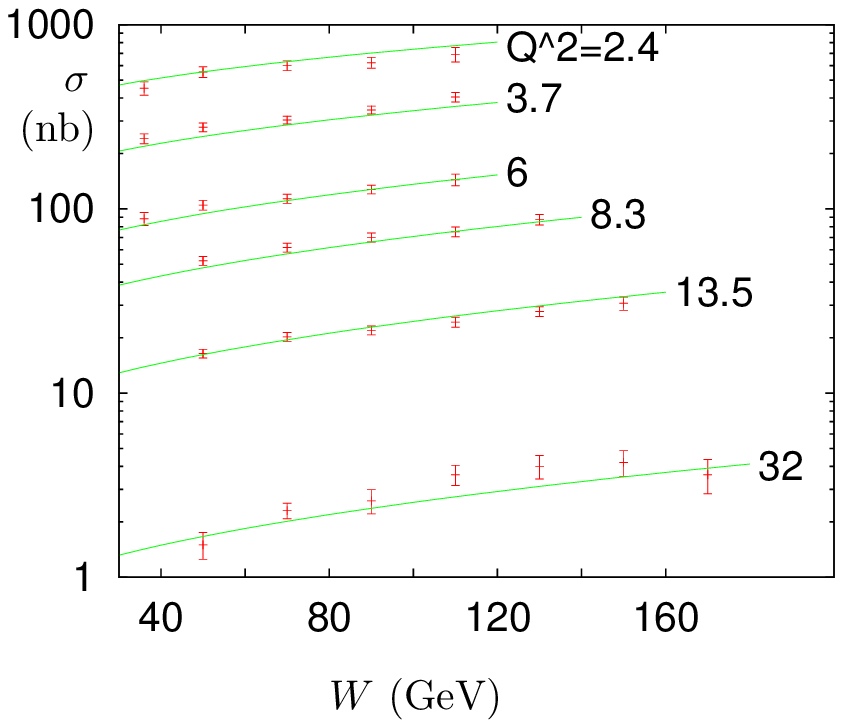}
\hfill
\epsfxsize=0.4\hsize\epsfbox[176 555 440 760]{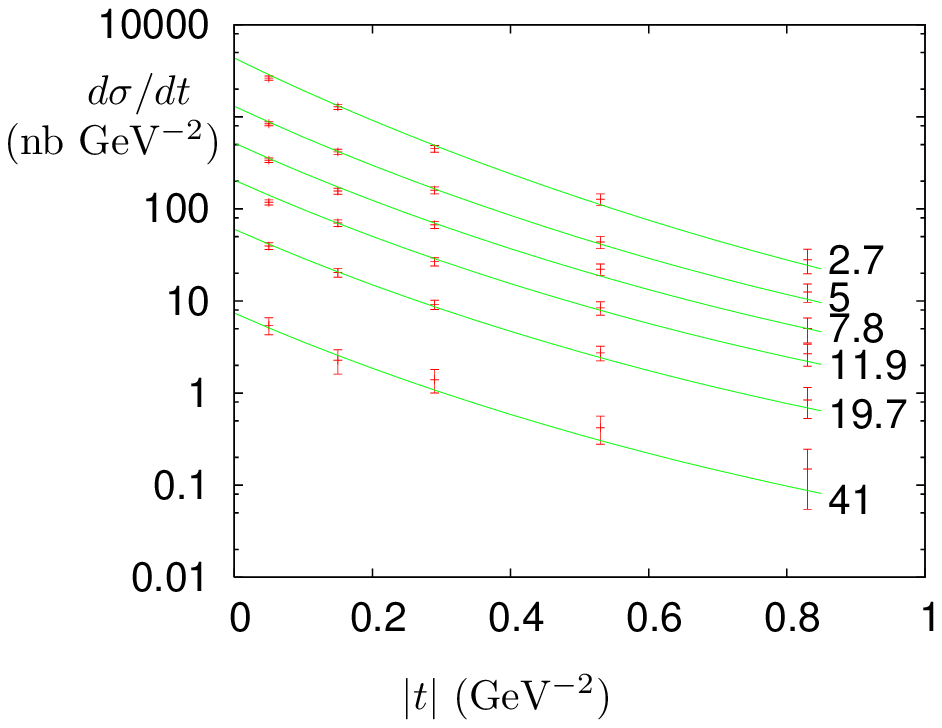}}

\line{\hfill (a) \hfill\hfill (b)\hfill}
Figure 3: Fits to exclusive $\rho$ electroproduction data\ref{\zeus}: 
(a) the total cross section and (b) the differential cross section at
$\langle W \rangle=90$~GeV, each at various values of $Q^2$.
\endinsert

{\bf Exclusive \pmb{$\phi$} electroproduction}

As for $\rho$ electroproduction, for the fit to $\phi$ electroproduction we 
consider only the hard and soft pomeron contributions and restrict the fit
to the high-energy data\ref{\phidat}. The data comprise differential cross 
sections at $\langle W \rangle = 75$ GeV at seven different values of $Q^2$, 
and the energy dependence at $\langle Q^2 \rangle = 5$~GeV$^2$ at three 
different values of $t$. The data were fitted independently at each value 
of $Q^2$ to determine $X_0$ and $X_1$ as functions of $Q^2$. We found 
that they could be well represented by
$$
X_0 = {1.28\over{(1+Q^2/Q_0^2)}}~~~~~~X_1 = {28.19\over{(1+Q^2/Q_1^2)^2}}
\eqno(6a)
$$
with
$$
Q_0^2 = 7.5~{\rm GeV}^2~~~~~~Q_1^2 = 5.0~{\rm GeV}^2,
\eqno(6b)
$$
except at $Q^2 = 5$ and 5.2 GeV$^2$. At 5(5.2) GeV$^2$ $X_0$ lies below(above) 
the curve and $X_1$ lies above(below). In (6a) the units are such that 
$|T(s,t)|^2$ is the differential cross section in nb GeV$^{-2}$. 
The data do not put a significant 
constraint on the mass $M$ of the form factor $G(t)$ of (3b) or on whether
separate form factors are required for the soft and hard pomeron, provided
only that $M$ is large for the hard pomeron. For simplicity a large value was 
taken for both so that $G(t)$ is effectively constant and the $t$ dependence
is given entirely by the Regge behaviour and the proton form factor. The fit
is shown in figure 4.

\topinsert
\line{\epsfxsize=0.4\hsize\epsfbox[176 555 440 760]{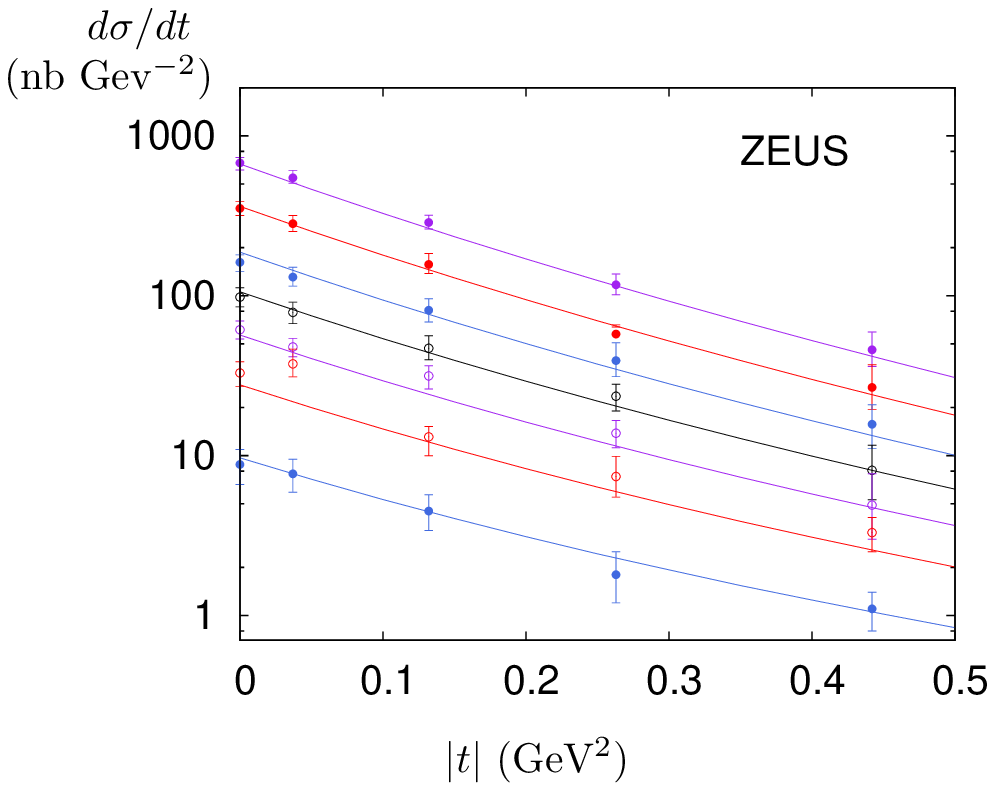}
\hfill
\epsfxsize=0.4\hsize\epsfbox[176 555 440 760]{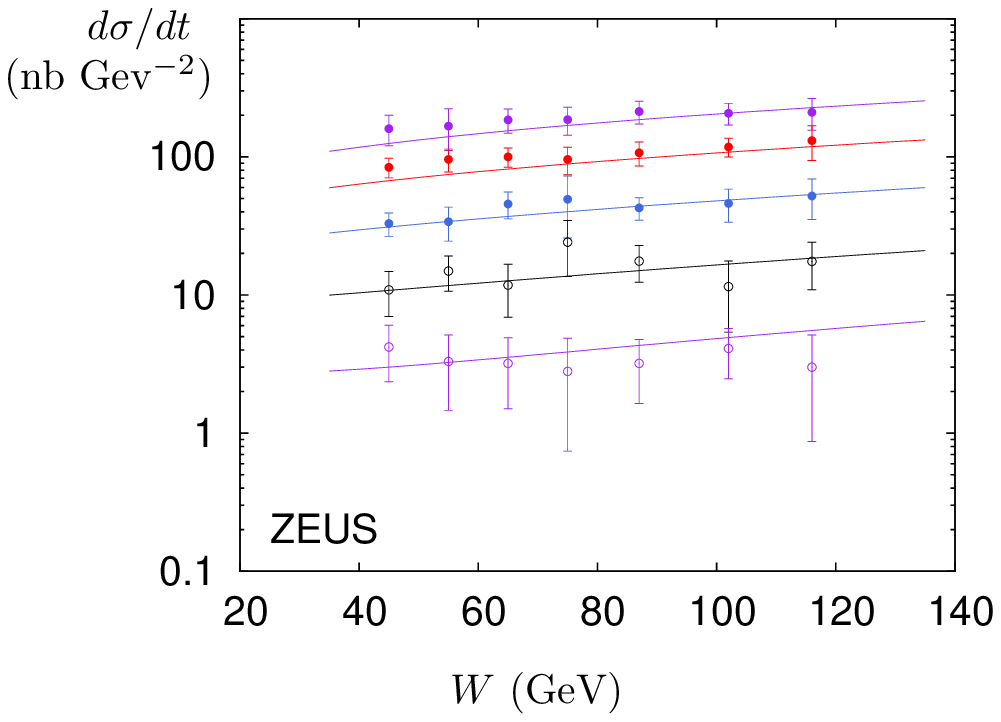}}

\line{\hfill (a) \hfill\hfill (b)\hfill}
Figure 4: Fit to exclusive $\phi$ electroproduction: (a) at $\langle W \rangle = 75$
GeV and $Q^2$ = 2.4, 3.6, 5.2, 6.9, 9.2, 12.6 and 19.7 GeV$^2$ 
and (b) at $\langle Q^2 \rangle = 5$ GeV$^2$ and $|t| =$ 0.025, 0.12, 0.25, 0.45
and 0.73 GeV$^2$. 
\endinsert

{\bf Exclusive \pmb{$J/\psi$} photoproduction}

For exclusive $J/\psi$ production from real photons, we use a fit
exactly similar to that for exclusive $\rho$ production. This results in
$$
X_0=1.78~~~~ X_1= 6.75~~~~ X_2=0.21~~~~ M=64.3~\hbox{GeV}
\eqno(7)
$$
where again the units are such that $|T(s,t)|^2$ is the differential cross
section in nb~GeV$^{-2}$.  This gives the fits shown in figure 5.

We have remarked previously that, while the soft pomeron seems not to couple
to $c$ quarks, it does couple to the $J/\psi$, and that there is old
evidence\defref\oldomega{
M J Corden et al, Physics Letters 68B (1977) 96
}
that the $J/\psi$ consists of a significant $q\bar q$ component, not just
$c\bar c$. The fit is not good at low energy and large $t$; we cannot explain
why the differential cross section is so large there.
\bigskip
\topinsert
\centerline{\epsfxsize=0.4\hsize\epsfbox[176 555 440 760]{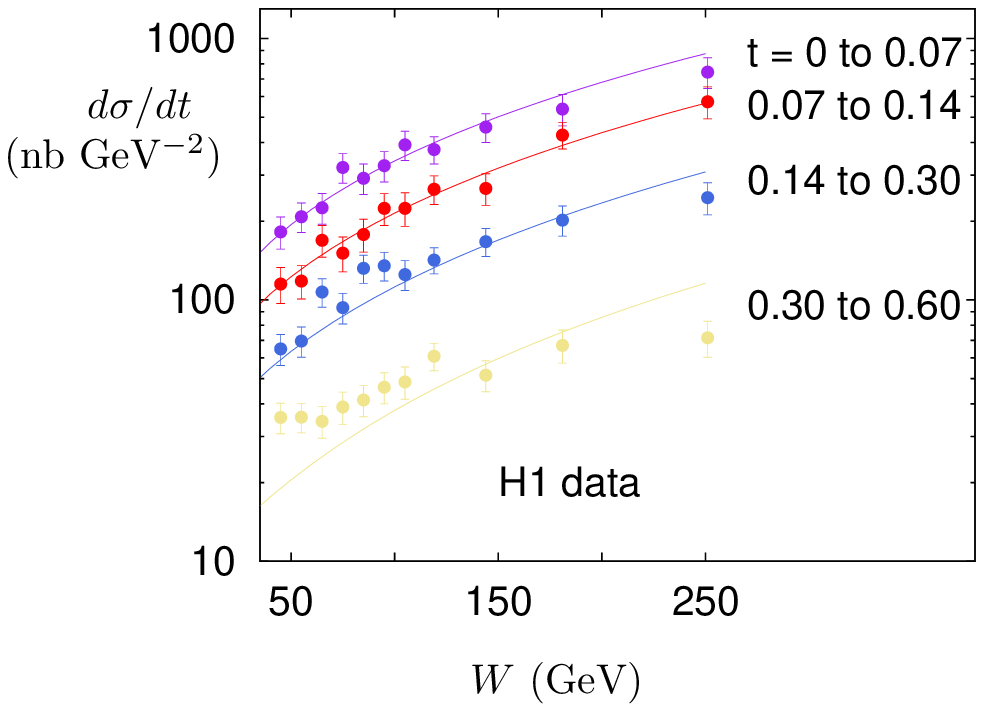}}
  
\line{\epsfxsize=0.4\hsize\epsfbox[176 555 440 760]{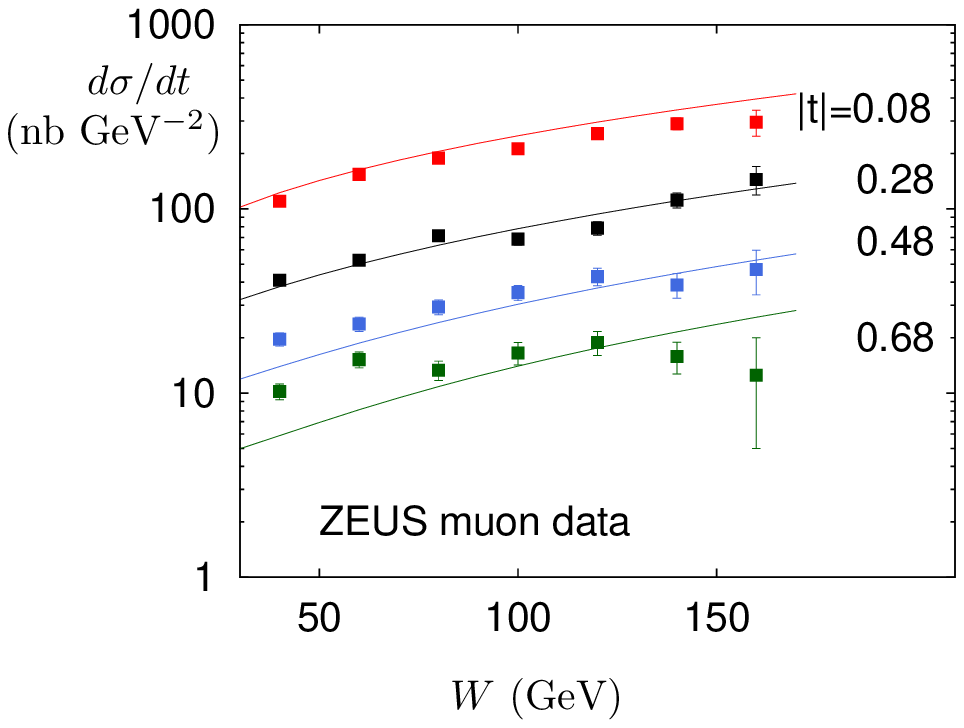}
\hfill
\epsfxsize=0.4\hsize\epsfbox[176 555 440 760]{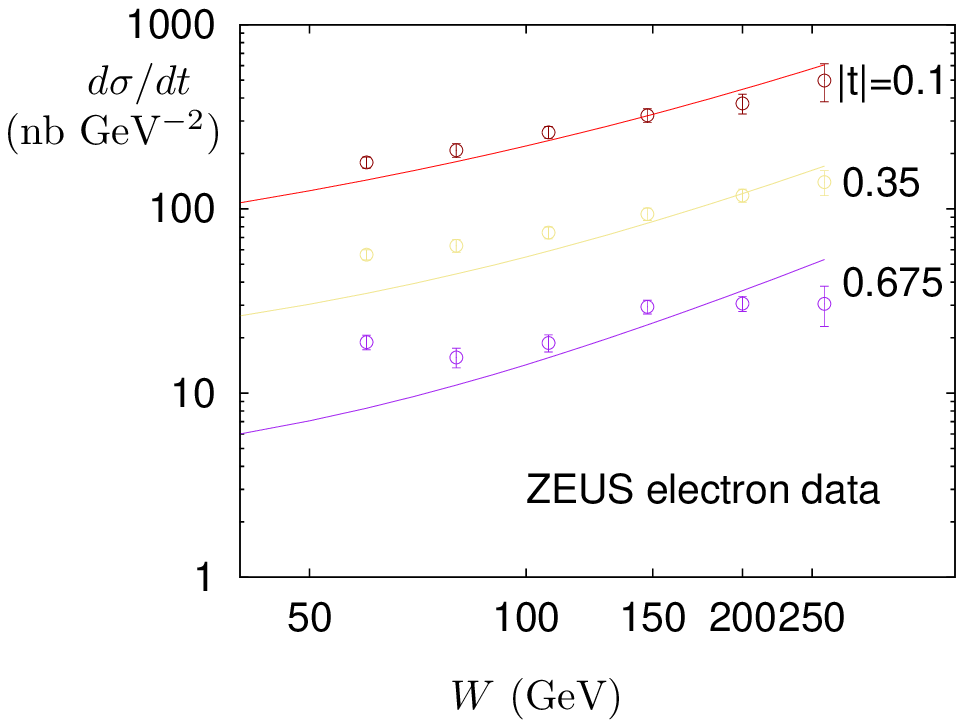}}

\centerline{Figure 5: Fits to data for exclusive $J/\psi$ photoproduction}
\bigskip
\bigskip
\bigskip
\centerline{\epsfxsize=0.4\hsize\epsfbox[176 555 440 760]{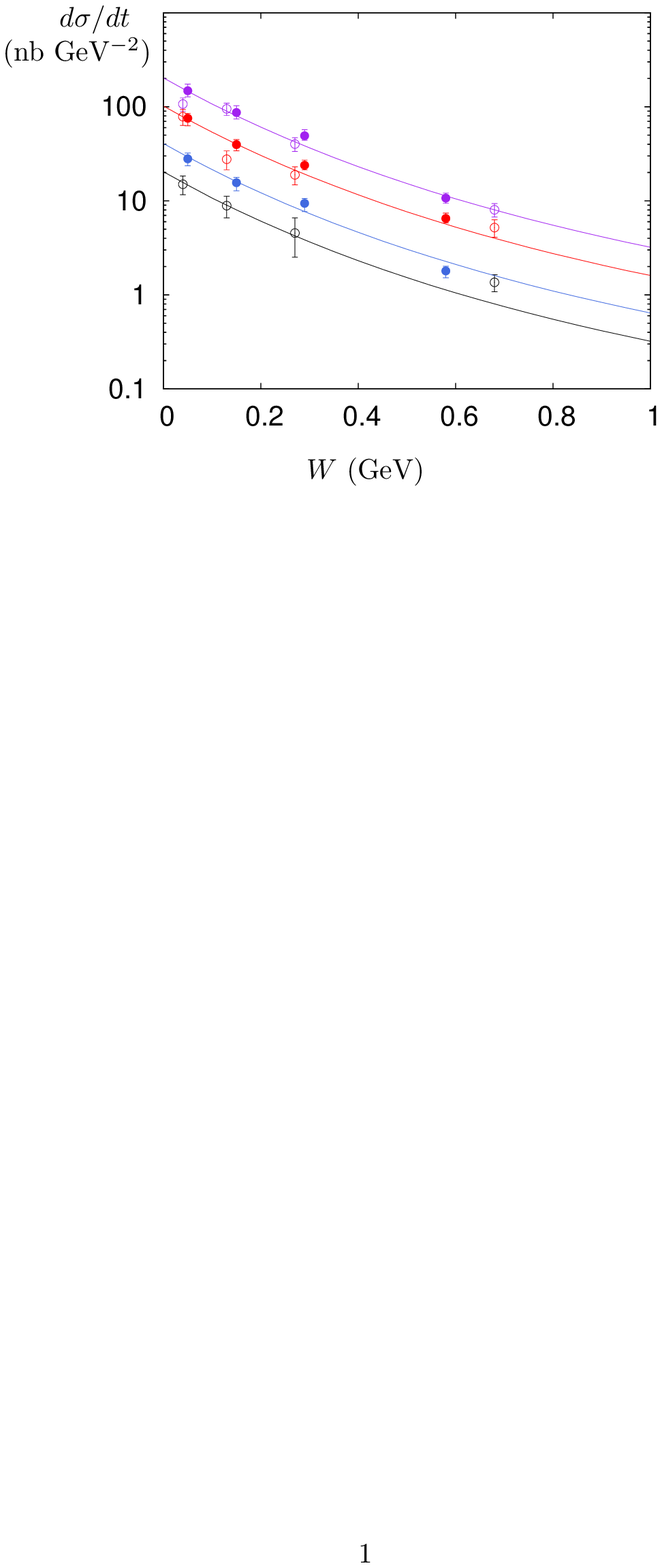}}

Figure 6: Fit to exclusive $J/\psi$ electroproduction differential
cross section at $\langle W \rangle=90$~GeV, at $Q^2 =$ 3.2, 7.0, 16 and
22.4 GeV$^2$.
 
\endinsert
{\bf Exclusive \pmb{$J/\psi$} electroproduction}

For exclusive $J/\psi$ electroproduction we again repeat the procedure used 
for $\rho$ electroproduction, first fitting the data at each $Q^2$ and then 
determining an appropriate $Q^2$ dependence. A feature of the $J/\psi$ 
electroproduction is that the energy dependence does not vary much with 
$Q^2$ or $t$ and is very close to that of small-$t$ $J/\psi$ photoproduction.
It is found adequate to give both $X_0$ and $X_1$ the same $Q^2$ dependence,
multiplying $X_0$ and $X_1$ of (7) by
$$
H(Q^2,Q_1^2)={1\over{(1+Q^2/Q_1^2)}}
\eqno(8a)
$$
with 
$$
Q_1^2 = 5.23~{\rm GeV}^2
\eqno(8b)
$$

The fit is shown in figure 6.
\bigskip
\bigskip
In conclusion, we have shown that Regge theory, with the inclusion of the hard 
pomeron, gives a good fit to data for the exclusive photo and electroproduction
of vector mesons. Thus we disagree with a view recently expressed\defref\jung{
H Jung, on behalf of the H1 and ZEUS collaborations, arXiv:0801.1970
}
that the understanding of elastic vector meson photoproduction is still a 
challenge.
\vfill\eject
\medskip\immediate\closeout\rfile\writestoppt
\baselineskip=14pt{{\bf References}}\bigskip{\frenchspacing%
\parindent=20pt\escapechar=` \input refs.tmp\bigskip}\nonfrenchspacing
\bye